\documentclass[twocolumn,showpacs,preprintnumbers,amssymb,prl]{revtex4}
\usepackage{graphicx}
\usepackage{dcolumn}
\usepackage{color}
\usepackage{bm}
\begin{document}

\title {\bf Broadcasting of continuous variable entanglement}
\author{Satyabrata Adhikari}
\altaffiliation{satyabrata@bose.res.in}
\affiliation{S. N. Bose National Centre for Basic Sciences,
Salt Lake, Kolkata 700 098, India}
\author{A. S. Majumdar}
\altaffiliation{archan@bose.res.in}
\affiliation{S. N. Bose National Centre for Basic Sciences,
Salt Lake, Kolkata 700 098, India}
\author{N. Nayak}
\altaffiliation{nayak@bose.res.in}
\affiliation{S. N. Bose National Centre for Basic Sciences,
Salt Lake, Kolkata 700 098, India}
\date{\today}

\vskip 0.5cm
\begin{abstract}
We present a scheme for broadcasting of continuous variable entanglement.
We show how an initial two-mode squeezed state of the electromagnetic 
field shared by two distant parties can be broadcasted into two nonlocal 
bipartite entangled states. Our protocol uses a local linear amplifier
and a beam splitter at each end. We compute the fidelity of the output 
entangled states and show that the broadcasting can be implemented for
a variety of input squeezed states and amplifier phases.
\end{abstract}

\pacs{03.67.Mn,42.50.Dv}

\maketitle

Quantum entanglement is now recognized as a powerful resource in 
communication and computation 
protocols\cite{nielsen}. The first nontrivial consequence of
entanglement on quantum ontology was noticed many years ago within the 
context of continuous variable systems\cite{EPR}. In recent times there has 
been a rapid development of the theory of entanglement pertaining to 
infinite dimensional Hilbert spaces\cite{reviews}. Many
well-known results of discrete variable systems relating to classification
and manipulation of entanglement take novel
forms for the case of continuous variables\cite{adesso}. There still remains 
a lot to be understood in the information theory for continuous variables
which has potentially vast practical ramifications. 

An interesting issue is that of broadcasting of quantum entanglement, viz.,
whether the entanglement shared by a single pair can be transmitted to two
less entangled pairs by local operations at both ends. Unlike classical
correlations, quantum entanglement cannot always be broadcasted, as has been
proved for general mixed states in finite dimensions\cite{barnum}. 
Since broadcasting involves 
copying of local information, and the exact cloning of an unknown quantum
state is impossible, the no-cloning theorem\cite{zurek} and its
consequences imply limitations on this procedure. For the case of pure states
in finite dimensions, implementation of broadcasting imposes restrictions
on the initial state\cite{buzek} and conditions on the fidelity of the 
cloning process\cite{kar}. No scheme has yet been 
proposed, however, for the broadcasting of continuous variable entanglement. 

The cloning of continuous quantum variables has nonetheless, been studied
by several authors. Various schemes for duplication of coherent states with 
optimal fidelity and economical means 
have been suggested\cite{cerf}. Operations of
cloning machines with networks of linear amplifiers and
beam splitters have been proposed\cite{braunstein,fiurasek}. 
It is thus relevant to investigate whether such ideas of copying
local information can be elaborated for broadcasting entangled states 
of continuous variables.
To this end we extend the procedure of cloning of a single-mode 
squeezed state of the electromagnetic field proposed by
Braunstein et al.\cite{braunstein} to the case of a bipartite entangled
two-mode squeezed state. By applying a linear amplifier and a beam splitter 
available locally with each party, we show using the covariance matrix 
approach\cite{adesso}  how the initial entangled state can be
broadcasted into two nonlocal and bipartite entangled states. 

Before describing our scheme for broadcasting in detail, 
it is instructive to review briefly
how the copying of local continuous variable information\cite{braunstein}
can be understood in terms of 
the covariance matrix approach. For this purpose, we
begin with a single-mode squeezed vacuum state of the electromagnetic
field represented by the squeezing transformation operator
\begin{eqnarray}
S_i(r) = \left(\begin{matrix}{{\mathrm e}^r & 0\cr 0 & {\mathrm e}^{-r}}\end{matrix}\right)
\label{singlemode1}
\end{eqnarray}
acting on the vacuum mode, and $r$ is the squeezing parameter ($r>0$).
The covariance matrix (CM) corresponding to the single-mode (say, $i$) 
squeezed vacuum state is given by
\begin{eqnarray}
\sigma_i(r) = S_i(r)S^T_i(r) = \left(\begin{matrix}{\mathrm{e}^{2r} & 0\cr
0 & \mathrm{e}^{-2r}}\end{matrix}\right)
\label{covmat0}
\end{eqnarray}

The cloning of this state proceeds as follows. First, a linear 
amplifier\cite{adesso} mediates the interaction
between the mode $i$ and an ancilla (say, $a$) prepared in the 
vacuum state, which is represented
by the linear transformation
\begin{eqnarray}
{\cal A}_{ia}(r,\phi) = [S_{ia}(r,\phi)].[S_i \oplus I_a] 
\label{amplifier}
\end{eqnarray}
with $\phi$ being the phase of the amplifier. After this interaction the
squeezed state mode together with the ancilla mode and another blank 
state mode (say $b$) are incident
on a three mode $50:50$ beam splitter $B_{iab}$ which we define through
a symplectic transformation as 
\begin{eqnarray}
B_{iab} = \left(\begin{matrix}{\sqrt{1/2} & 0 & 0 & 0 & \sqrt{1/2} & 0\cr
0 & \sqrt{1/2} & 0 & 0 & 0 & \sqrt{1/2}\cr 0 & 0 & 1 & 0 & 0 & 0\cr
0 & 0 & 0 & 1 & 0 & 0\cr \sqrt{1/2} & 0 & 0 & 0 & -\sqrt{1/2} & 0\cr
0 & \sqrt{1/2} & 0 & 0 & 0 & -\sqrt{1/2}}\end{matrix}\right)
\label{beamsplit2}
\end{eqnarray}
The above beam splitter is defined in such a way that it does not
affect the ancilla mode.
The total cloning operation is thus represented by the transformation
\begin{eqnarray}
T_{iab}(r,\phi) = [B_{iab}].[{\cal A}_{ia} \oplus I_b]
\label{clone}
\end{eqnarray}
with the corresponding CM given by
\begin{eqnarray}
\sigma_{iab} = T_{iab}(r,\phi)T^{\dagger}_{iab}(r,\phi)
\label{covmat01}
\end{eqnarray}
This procedure leads to symmetric cloning resulting in the two cloned output
modes $i$ and $b$. 

The fidelity of the two clones can be evaluated 
through the relation\cite{fidelity}
\begin{eqnarray}
F = \frac{1}{\sqrt{\mathrm{Det}[\sigma_{in} + \sigma_{out}]+\delta}-\sqrt{\delta}}
\label{fidel0}
\end{eqnarray}
where $\sigma_{in}$ is given by Eq.(\ref{covmat0}), and $\sigma_{out}$ is
obtained by tracing out the ancilla mode from the CM in Eq.(\ref{covmat01}), 
i.e.,
\begin{eqnarray}
\sigma_{out} = \left(\begin{matrix}{P & 0\cr 0 & M}\end{matrix}\right)
\label{clone0}
\end{eqnarray}
with
\begin{eqnarray}
P \hskip -0.2cm &=& \hskip -0.1cm \frac{{\mathrm e}^{2r}(c-hs)^2+k^2s^2+1}{2}, \hskip 0.1cm
M \hskip -0.1cm = \frac{{\mathrm e}^{-2r}(c+hs)^2+k^2s^2+1}{2}\nonumber\\
\delta &=& 4(\mathrm{Det}[\sigma_{in}]-1/4)(\mathrm{Det}[\sigma_{out}]-1/4)
\label{defs0}
\end{eqnarray}
where
\begin{eqnarray}
c \hskip -0.05cm = \mathrm{Cosh}(2r), \hskip 0.1cm s \hskip -0.05cm = \mathrm{Sinh}(2r), \hskip 0.1cm 
h \hskip -0.05cm = \mathrm{Cos}(2\phi), \hskip 0.1cm k \hskip -0.05cm = \mathrm{Sin}(2\phi)
\label{defs002}
\end{eqnarray}
The fidelity for the above phase sensitive cloning machine is thus given by
\begin{eqnarray}
F \hskip -0.2cm = \hskip -0.2cm \frac{1}{\sqrt{\hskip -0.05cm(P \hskip -0.1cm +  \mathrm{e}^{2r})\hskip -0.05cm(M \hskip -0.1cm +\mathrm{e}^{-2r})\hskip -0.1cm + \hskip -0.1cm 3(PM \hskip -0.1cm - \hskip -0.1cm 1/4)} 
\hskip -0.1cm - \hskip -0.2cm \sqrt{3(PM \hskip -0.1cm - \hskip -0.1cm 1/4)}}
\label{fidel1}
\end{eqnarray}
If the phase of the 
amplifier is set to $\phi=0$, the fidelity becomes
\begin{eqnarray}
F=\frac{2}{\sqrt{8c^2+12c+5}-\sqrt{3+6c}}
\label{fidel2} 
\end{eqnarray}
Since the fidelity of the clones depend on the input state, the cloning is
said to be state-dependent. From Eq.(\ref{fidel2}) it follows that as
$r \to \infty$, $F \to 0$, and as $r \to 0$, $F \to 1$.

Let us now consider a continuous variable entangled state 
(an entangled state of the electromagnetic field) which is shared by two
parties located far apart at sites I and J, respectively, and is 
represented by the generic two-mode ($i$ 
and $j$) squeezed state with one mode at each end. 
The two-mode squeezed vacuum state is obtained by applying
the transformation\cite{adesso}
\begin{eqnarray}
T_{ij}(r) = B_{ij}(1/2).(S_i(r)\oplus S_j(r))
\label{twomode1}
\end{eqnarray}
on two uncorrelated squeezed vacuum modes ($S_i(r)$ and $S_j(r)$) given by 
Eq.(\ref{singlemode1}). $B_{ij}(1/2)$ denotes a balanced
$50:50$ beam splitter with the matrix form
\begin{eqnarray}
B_{ij}(1/2) = \left(\begin{matrix}{\sqrt{1/2}& 0 & \sqrt{1/2} & 0\cr
0 & \sqrt{1/2} & 0 & \sqrt{1/2}\cr \sqrt{1/2} & 0 & -\sqrt{1/2} & 0\cr
0 & \sqrt{1/2} & 0 & -\sqrt{1/2}}\end{matrix}\right)
\label{beamsplit1}
\end{eqnarray}
The CM corresponding to the
two-mode squeezed state is given by
\begin{eqnarray}
\sigma_{ij}(r) = T_{ij}(r)T^{\dagger}_{ij}(r) =\left(\begin{matrix}
{c &0&s&0\cr 0&c&0&-s\cr s & 0 & c & 0\cr 0 & -s & 0 & c}\end{matrix}\right)
\label{twomode2}
\end{eqnarray}  
The two-mode squeezed vacuum state is the quantum optical representative for
bipartite continuous variable entanglement.
In the Heisenberg picture
the quadrature operators of the two-mode squeezed state are given by
\begin{eqnarray}
\hat{x}_i = \frac{{\mathrm e}^r \hat{x}_i^{(0)} + {\mathrm e}^{-r}\hat{x}_j^{(0)}}{\sqrt{2}}, \hskip 0.3cm 
\hat{p}_i = \frac{{\mathrm e}^{-r} \hat{p}_i^{(0)} + {\mathrm e}^r\hat{p}_j^{(0)}}{\sqrt{2}}\nonumber\\
\hat{x}_j = \frac{{\mathrm e}^r \hat{x}_i^{(0)} - {\mathrm e}^{-r}\hat{x}_j^{(0)}}{\sqrt{2}}, \hskip 0.3cm 
\hat{p}_j = \frac{{\mathrm e}^{-r} \hat{p}_i^{(0)} - {\mathrm e}^r\hat{p}_j^{(0)}}{\sqrt{2}}
\label{quad}
\end{eqnarray}
where the superscript $(0)$ denotes the initial vacuum modes, the operators
$\hat{x}$  and $\hat{p}$ represent the electric quadrature amplitudes (the 
real and imaginary parts, respectively, of the mode's annihilation operator).

Now, for broadcasting of the above two-mode squeezed state we apply local
cloning machines on the two
individual modes of the entangled bipartite state,
located at the sites I and J, respectively.
Our scheme for broadcasting proceeds as follows. The local cloner
acting on the mode at site I copies the information available locally
on to two modes ($i$
and $b$). Similarly, the cloner acting on the mode at site J copies information
on to two modes ($j$ and $b'$). Note that since the two modes on 
which the cloners act at 
sites I and J, respectively,  are the constituents of an initially 
entangled bipartite state,
the forms of the output local clones will be different in general, from
the outputs of the cloning for a single-mode squeezed state given 
by Eq.(\ref{clone0}). More importantly, the properties of entanglement 
between
the output states are now dependent on the initial entangled state to
be broadcasted. 

The criteria for successful
broadcasting\cite{buzek,kar} of the entangled two-mode squeezed 
state (\ref{twomode2}) 
can be elaborated as follows. The initial
entangled state is broadcasted if the local pairs of output modes ($i$ and $b$ 
at site I) and ($j$ and $b'$ at site J) are separable, and also if 
simultaneously, both the non-local pairs of output modes 
($i$ and $b'$ on one hand),
and ($j$ and $b$ on the other) are entangled. 
Our aim is to verify
whether the above conditions are satisfied for the output states when the
two cloners act at their respective sites. To this end, we formulate
this bi-local cloning procedure  using the CM approach\cite{adesso}. 

The broadcasting operation is implemented through an ancilla mode, a linear
amplifier and a beam splitter located at both sites. Thus, after introducing 
the ancillas $a$ and $a'$ (at the ends I and J, respectively), the CM of
the joint two-mode squeezed state with the ancillas takes the form
\begin{eqnarray}
\sigma_{iaja'}(r) = \left(\begin{matrix}{c & 0 & 0 & 0 & s & 0 & 0 & 0\cr
0 & c & 0 & 0 & 0 & -s & 0 & 0\cr 0 & 0 & 1 & 0 & 0 & 0 & 0 & 0\cr
0 & 0 & 0 & 1 & 0 & 0 & 0 & 0\cr s & 0 & 0 & 0 & c & 0 & 0 & 0\cr
0 & -s & 0 & 0 & 0 & c & 0 & 0\cr 0 & 0 & 0 & 0 & 0 & 0 & 1 & 0\cr
0 & 0 & 0 & 0 & 0 & 0 & 0 & 1}\end{matrix}\right)
\label{twoancilla}
\end{eqnarray}
Next, both parties apply linear amplifiers on their 
respective modes and ancillas. The local amplifiers can be jointly represented
as
\begin{eqnarray}
{\cal A}(r,\phi) = \left(\begin{matrix}{{\cal A}_1 & 0\cr 0 & {\cal A}_2}\end{matrix}\right)
\label{twoamplf}
\end{eqnarray}
where the ${\cal A}_i$ are given by\cite{adesso}
\begin{eqnarray}
{\cal A}_i = \left(\begin{matrix}{c-hs & 0 & ks & 0\cr 0 & c+hs & 0 & -ks\cr
ks & 0 & c+hs & 0\cr 0 & -ks & 0 & c-hs}\end{matrix}\right)
\label{twoamplf2}
\end{eqnarray} 
for $i=1,2$. After interaction with the local amplifiers, the CM of
the two modes with their ancillas is transformed to 
\begin{eqnarray}
\sigma'_{iaja'}(r,\phi) =
{\cal A}^{T}\sigma_{iaja'}(r,\phi){\cal A}. 
\end{eqnarray}

Thereafter, both parties introduce their
respective blank modes ($b$ and $b'$) on which the information of the 
original modes is to
be copied. Both parties now have three local modes each, i.e., original,
ancilla, and blank mode, which fall on a $50:50$ beam splitter 
defined through a symplectic transformation in Eq.(\ref{beamsplit2}), 
at each end. The two local beam splitters can
be represented jointly by 
\begin{eqnarray}
B = \left(\begin{matrix}{B_{iab} & 0\cr 0 & B_{ja'b'}}\end{matrix}\right)
\label{beamsplit3}
\end{eqnarray}
and the resultant CM at the end of the cloning processes at both
the ends is given by 
\begin{eqnarray}
\sigma''_{iabja'b'} = B^{T}\sigma'_{iabja'b'}B
\end{eqnarray} 

Using the above CM we can now check if the criteria for
successful broadcasting are satisfied. In order to verify the entanglement of
the non-local pairs of modes shared by the two sides, we 
obtain the reduced CMs corresponding to these modes,
which (after tracing out the remaining modes from $\sigma''_{iabja'b'}$) 
are given by
\begin{eqnarray}
\sigma^{\mathrm{non-local}}_{ib'}(r,\phi) &=& \sigma^{\mathrm{non-local}}_{jb}(r,\phi)\nonumber\\ &=& \left(\begin{matrix}{\frac{G+1}{2} & 0 & \frac{E}{2} & 0\cr 0 &
\frac{H+1}{2} & 0 & \frac{-E}{2}\cr
\frac{E}{2}& 0 & \frac{G+1}{2} & 0\cr 0 & \frac{-E}{2} & 0 & \frac{H+1}{2}}\end{matrix}\right)
\label{covmat2}
\end{eqnarray}
where 
\begin{eqnarray}
E \hskip -0.1cm = s(c-hs)^2 \hskip -0.05cm , \hskip 0.1cm G \hskip -0.1cm = (c-hs)^2c \hskip -0.05cm + \hskip -0.05cm ks^2 \hskip -0.05cm , \nonumber\\
H \hskip -0.1cm = (c+hs)^2c \hskip -0.05cm + \hskip -0.05cm ks^2
\label{defs3}
\end{eqnarray}
The condition for separability of the modes can be 
obtained from the
generalyzation of the positivity of partial transposition (PPT) criterion
for continuous variable systems\cite{simon}. For a two-mode state represented
by the CM (\ref{covmat2}), the necessary and sufficient condition for the
separability of these modes\cite{adesso} reduces to the relation $-E^2/4 >0$
being satisfied. Since the above inequality is always violated, it follows
that the non-local output modes are
always entangled. 

\begin{figure}[h!]
\begin{center}
\includegraphics[width=10cm]{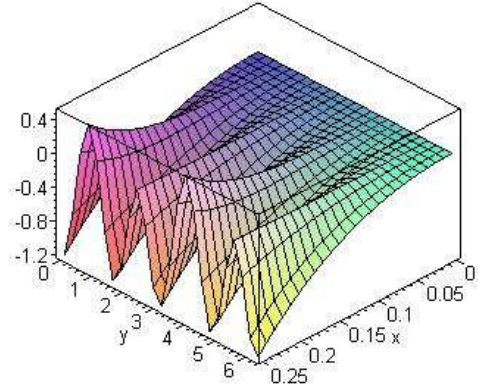}
\vskip -0.5cm
\caption{The separability parameter $R$ is plotted versus the 
squeezing $r$ ($x$-axis), and the amplifier phase $\phi$ ($y$-axis). The 
condition
for separabilty of the local modes ($R>0$) is satisfied for various
combinations of $r$ and $\phi$.}
\end{center}
\label{f1}
\end{figure}

\vskip -0.5cm

Our remaining task for implementing broadcasting is to find the conditions
under which the local output modes turn out to be separable. 
After tracing out the ancilla 
mode on site I and also all the modes on the site
J from $\sigma''_{iabja'b'}$, the reduced CM representing the
system of the two clones on site I (which is also equal to the corresponding
reduced CM on the site J) is given by
\begin{eqnarray}
\sigma^{\mathrm{local}}_{ib}(r,\phi) &=& \sigma^{\mathrm{local}}_{jb'}(r,\phi)\nonumber\\ &=& \left(\begin{matrix}{\frac{G+1}{2} & 0 & \frac{G-1}{2} & 0\cr 0 &
\frac{H+1}{2} & 0 & \frac{H-1}{2}\cr
\frac{G-1}{2}& 0 & \frac{G+1}{2} & 0\cr 0 & \frac{H-1}{2} & 0 & \frac{H+1}{2}}\end{matrix}\right)
\label{covmat1}
\end{eqnarray}
The PPT condition\cite{simon} for the local output modes 
represented
by the CM (\ref{covmat1}) is satisfied\cite{adesso} if the relation 
\begin{eqnarray}
\hskip -0.2cm R \hskip -0.1cm = \hskip -0.1cm [(c-hs)^2c+(ks^2-1)][(c+hs^2)c+(ks^2-1)]\hskip -0.1cm > \hskip -0.1cm 0
\label{cond1}
\end{eqnarray}
holds. It follows 
from Eq.(\ref{cond1}) that whenever $ks^2 > 1$, the output
local clone modes are separable.
We display in Fig.1, the three-dimensional plot of the variable $R$
as a function of the squeezing parameter $r$ and the amplifier phase $\phi$.
One sees that this criterion for broadcasting can be satisfied for 
several values of the above two parameters. The pattern of $R$ versus $\phi$
continues for large values of squeezing too. For example, if the phase of
the amplifiers is set to $\phi = \pi/4$, then broadcasting is possible 
for all values of the squeezing parameter $r$, i.e., for all continuous
variable bipartite entangled states that are represented by the two-mode
squeezed state (\ref{twomode2}).

\begin{figure}[h!]
\begin{center}
\includegraphics[width=9cm]{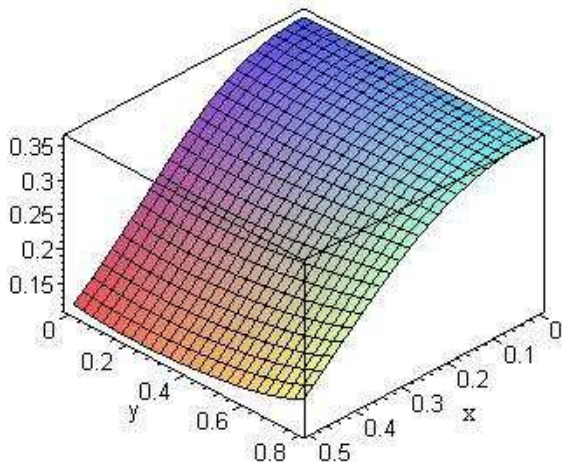}
\caption{The fidelity of broadcasting $F_B$ is plotted versus the 
squeezing $r$ ($x$-axis), and the amplifier phase 
$\phi$ ($y$-axis).}
\end{center}
\label{f2}
\end{figure}

\vskip -0.5cm

We have shown that broadcasting is possible in our scheme for various
combinations of input states and amplifier phases. Finally, one would
like to investigate the efficiency of broadcasting through this scheme.
For this purpose we compute the fidelity of the broadcasted states, i.e.,
the entangled non-local states (\ref{covmat2}). The fidelity of broadcasting,
$F_B$ is obtained through Eq.(\ref{fidel0}), in which we substitute the
expressions for $\sigma_{in}$ and $\sigma_{out}$ from Eqs.(\ref{twomode2})
and (\ref{covmat2}) respectively. In Fig.2 we display $F_B$ as 
a function of the 
squeezing parameter $r$ and the amplifier phase $\phi$. One sees that this
scheme of ours yields a phase- and state-dependent broadcasting fidelity. From
the obtained expression for $F_B$ it is possible to see that 
for $r \to \infty$, $F_B \to 0$, and for $r \to 0$, $F_B \to 0.36$.

To summarize, in this paper we have presented  a scheme for broadcasting
of continuous variable entanglement for the first time. We consider an
initial two-mode squeezed state which is a generic bipartite entangled
state of the electromagnetic field. This state is shared by two distant
parties who individually apply  local cloning machines on their respective
modes. The cloning process\cite{braunstein} at each end requires an ancilla
state, a linear amplifier and a beam splitter, yielding two symmetric cloned 
modes each. The initial
state is broadcasted into a pair of bipartite entangled 
states that is finally shared by the two distant parties.
We show using the covariance matrix formalism\cite{adesso} 
when the ouput states with the two distant parties
satisfy the criteria required for successful broadcasting.  We also compute the
fidelity of broadcasting using this procedure. Though our protocol for
broadcasting relies on phase sensitive and state-dependent cloners,
we find that it is successful for various combinations
of the squeezing parameter and the phase of the amplifiers. 


We conclude by noting some possible off-shoots of our present study.
It may be interesting to investigate other cloning protocols for continuous 
variables\cite{cerf,fiurasek} in order 
to see if they could lead to more efficient broadcasting. Whether broadcasting
is possible for mixed states in general\cite{barnum}, is itself an open 
question for infinite dimensional systems. Furthermore, the possibility of
generating three-mode quantum channels through local operations\cite{satya1} 
useful for communications
could be explored with photons. Finally, it may be noted that the first
experimental demonstration of continuous variable cloning has been reported 
recently\cite{subuncu}, and with further development it could be feasible
to experimentally broadcast entangled states of continuous variables too.

\end{document}